# Deep Learning-Driven Protein Structure Prediction and Design: Key Model Developments by Nobel Laureates and Multi-Domain Applications

Wanqing Yang[1,2], Yanwei Wang[1,2#] and Yang Wang[2#],

1. College of Mathematics and Physics, Wenzhou University, Wenzhou, 325000, China
2. Wenzhou Key Laboratory of Biomedical Imaging, Center of Biomedical Physics, Wenzhou Institute, University of Chinese Academy of Sciences, Wenzhou, 325000, China

*Corresponding author:

wangyw@wzu.edu.cn    and    wangy0727@ucas.ac.cn

# Abstract

This systematic review outlines pivotal advancements in deep learning-driven protein structure prediction and design, focusing on four core models—AlphaFold, RoseTTAFold, RFDiffusion, and ProteinMPNN—developed by 2024 Nobel Laureates in Chemistry: David Baker, Demis Hassabis, and John Jumper. We analyze their technological iterations and collaborative design paradigms, emphasizing breakthroughs in atomic-level structural accuracy, functional protein engineering, and multi-component biomolecular interaction modeling. Key innovations include AlphaFold3's diffusion-based framework for unified biomolecular prediction, RoseTTAFold's three-track architecture integrating sequence and spatial constraints, RFDiffusion's denoising diffusion for de novo protein generation, and ProteinMPNN's inverse folding for sequence-structure co-optimization. Despite transformative progress in applications such as binder design, nanomaterials, and enzyme engineering, challenges persist in dynamic conformational sampling, multimodal data integration, and generalization to non-canonical targets. We propose future directions, including hybrid physics-AI frameworks and multimodal learning, to bridge gaps between computational design and functional validation in cellular environments.

# Introduction

Proteins, as the central executors of life activities, derive their functional diversity from the precise folding of complex three-dimensional structures and dynamic regulation. For decades, elucidating structure-function relationships and achieving rational protein design has remained a core challenge in structural and synthetic biology. Traditional methods relying on X-ray crystallography and nuclear magnetic resonance (NMR) to resolve static conformations face limitations such as low experimental throughput and difficulties in capturing dynamic information, hindering systematic functional design. This impasse has been revolutionized by deep learning: data-driven neural network approaches not only overcome computational complexity constraints of traditional physical modeling but also establish a novel "predict-design-validate" methodology for protein engineering by deciphering evolutionary covariation signals and geometric constraints.

In recent years, AI-driven protein research has achieved landmark breakthroughs. The 2024 Nobel Prize in Chemistry was awarded to three pioneers in this field—David Baker from the Institute for Protein Design at the University of Washington, and John Jumper and Demis Hassabis from DeepMind—marking the transition of deep learning from an auxiliary tool to a transformative force. DeepMind's AlphaFold series, integrating attention mechanisms and geometric equivariant networks, achieves atomic-level accuracy in structure prediction. David Baker, who developed the initial version of Rosetta in 1998 and designed the first de novo non-natural protein in 2003, has recently pioneered AI-enhanced tools including RoseTTAFold, RFDiffusion, and ProteinMPNN (combining diffusion models with inverse folding algorithms). These innovations enable end-to-end automated design from topological scaffold generation to functional sequence optimization. As it is shown in Figure 1, the synergy between these two technological paradigms not only fills theoretical gaps in "sequence-structure-function" mapping but also catalyzes groundbreaking applications such as artificial enzymes, nanomaterials, and high-affinity binders.

Deep learning technology simulates the operational mechanisms of the human brain's neural networks by constructing multi-layered neural network architectures. Through self-attention mechanisms and geometric equivariant networks, it extracts latent information from massive evolutionary and structural databases to learn nonlinear sequence-structure-function relationships. This approach transcends the limitations of traditional physical modeling and empirical force fields, offering superior efficiency and accuracy in complex scenarios. This review summarizes the significant contributions of Nobel laureates in recent years, emphasizing core model innovations and their synergistic effects in function-oriented design. Furthermore, it discusses challenges such as multimodal data integration and dynamic conformational modeling, proposing a theoretical framework for developing next-generation protein design platforms that balance interpretability and creativity.

# 1 Prediction and Generation of Protein Structures

## 1.1 Iterative Innovations in the AlphaFold Series

AlphaFold, a revolutionary protein structure prediction system developed by DeepMind, has driven transformative advancements in structural biology through iterative architectural upgrades to its algorithmic framework. Since the debut of its first-generation model in 2018, this series has progressively overcome accuracy and applicability constraints in protein structure prediction via innovations in multi-scale modeling strategies and dynamic feature extraction mechanisms.

The original AlphaFold employed a convolutional neural network (CNN) [1] to establish a multi-task learning framework. It integrated homologous sequence alignments, torsional dihedral angle prediction, and residue distance distribution modeling, optimizing predicted structures through energy minimization strategies. During the CASP13 evaluation, the model achieved atomic-level accuracy predictions for select test proteins, demonstrating significantly reduced root mean square deviation (RMSD) compared to traditional homology modeling approaches. However, its heavy reliance on high-precision physical modeling incurred substantial computational resource demands, while increased prediction errors for proteins with low sequence homology revealed limitations in the recognition accuracy of critical structural features.

In 2020, DeepMind published the revolutionary AlphaFold2 model in *Nature*[2],[3]. Breaking free from the limitations of its CNN-based predecessor, this iteration introduced a dual-module system grounded in attention mechanisms: the Evoformer feature evolution module and the Structure Module, achieving end-to-end protein structure prediction for the first time. The Evoformer processes dual-track information flows—multiple sequence alignments (MSA) and residue pair representations—through 48 stacked transformer blocks. It resolves challenges in modeling long-range interactions and predicting complex topological folds by iteratively exchanging information while embedding physical constraints such as torsional angle restrictions and spatial distance thresholds. The Structure Module employs an SE(3)-equivariant neural network, synchronizing updates to local bond angles and global 3D coordinates via multi-scale optimization, enabling atomic-level spatial mapping from abstract features[4]. At CASP14, AlphaFold2 achieved a Global Distance Test (GDT_TS) score exceeding 90 for 87.4% of test proteins, with Cα atom RMSD below 2 Å for challenging targets like T1074, rivaling cryo-EM resolution[5]. Notably, it maintained atomic-level accuracy for core structural domains even under low-sequence-homology conditions, marking a qualitative leap in prediction capabilities. In 2021, DeepMind extended this framework to protein complexes with AlphaFold-Multimer[6]. The model integrates evolutionary signals from different subunits through MSA feature concatenation and enhances interface feature extraction using interface-specific attention mechanisms. Its loss function incorporates geometric constraints for binding interfaces, while a dynamic chain pruning strategy optimizes conformational sampling, balancing global complex topology and local interface precision. Systematic benchmarking revealed 67% overall accuracy (5Å distance threshold) for heteromeric interfaces, with 23% high-confidence predictions (pLDDT > 90). For homomeric complexes, accuracy reached 69% overall and 34% high confidence. Remarkably, the model successfully identified intricate folds like TIM barrels, demonstrating interface characterization capabilities critical for rational drug design and protein engineering.

The AlphaFold3 system, released in 2024, achieved a paradigm breakthrough in biomolecular interaction prediction[7],[8]. As it is shown in Figure 2, this model replaces the Evoformer with a Pairformer module, reducing computational complexity by 38% through decreased reliance on Multiple Sequence Alignment (MSA). Furthermore, it introduces a novel diffusion module that supersedes the original structural module, circumventing complex structural parameterization processes. The diffusion module samples atomic coordinates directly from noise distributions through a reverse denoising process, enabling the handling of complex molecular interactions without requiring predefined rotational frames. As the first unified multi-component biomolecular modeling framework, AlphaFold3 achieves full-atom collaborative prediction of various interactions, including protein-nucleic acid and protein-small molecule complexes. Evaluations across multiple benchmark datasets (Pose Busters, Recent PDB evaluation sets, and CASP15 RNA) demonstrate its superior performance over existing interface-specific methods in nearly all categories.

The AlphaFold technology series has achieved unprecedented milestones to date. Its predictive outcomes not only facilitate core biomedical research areas such as drug target identification[9]–[11]and rational enzyme design, but also drive paradigm innovations in dynamic analysis of complex protein assemblies and mechanistic studies of disease-related mutation[12]–[14]. Recent advancements have extended its applications to RNA secondary structure prediction[15]. The successful implementation of AlphaFold has generated massive predictive datasets. The AlphaFold Protein Structure Database (AlphaFold DB[16] initially encompassed over 360,000 predicted structures across 21 model organisms. Following its substantial 2024 upgrade, the database significantly enhanced structural feature mining efficiency through optimized multimodal data interfaces and interactive visualization tools, laying the foundation for realizing the ambitious goal of covering hundreds of millions of protein structures [17].

**1.2 Three-Track Architecture of RoseTTAFold**

In 2018, David Baker's team participated in CASP 13 using Rosetta[18], a protein design software based on energy function optimization and classical computational methods. Inspired by technical comparisons with the champion model AlphaFold, the research team spent three years refining their algorithms. It successfully developed RoseTTAFold[19] in 2021—a deep learning-based platform for protein structure and interaction prediction. During CASP 14, RoseTTAFold achieved groundbreaking performance: its global structure prediction accuracy (GDT_TS score) for distantly homologous proteins (sequence identity <25%) reached approximately 70, representing a ~42% improvement over the RosettaCM method.

As it is shown in Figure 3, RoseTTAfold employs a three-track neural architecture that synergistically processes and integrates protein features through parallel information channels: amino acid sequence characteristics (1D), spatial relationships between residues (2D), and three-dimensional coordinate information (3D). Cross-scale information exchange across tracks is achieved via gated attention mechanisms, with the 3D track utilizing SE(3)-Transformer networks[20]to directly optimize atomic coordinates, overcoming limitations of traditional methods reliant on 2D contact maps for structural prediction. When provided with target sequence data and ligand constraints, the model predicts residue contact maps by capturing evolutionary correlations between amino acids, directly generating 3D atomic coordinates. This end-to-end collaborative optimization mechanism enables RoseTTAFold to incorporate physical modeling constraints (e.g., van der Waals contacts, hydrogen bonding networks) into the deep learning process, enhancing the physical plausibility of predictions while maintaining data-driven efficiency.

Compared to purely data-driven prediction models, RoseTTAFold inherits Rosetta's core strengths: global exploration of conformational space via Monte Carlo simulations combined with molecular force field-based energy functions (including van der Waals potentials, electrostatic forces, and solvation effects) for local refinement. This hybrid strategy effectively mitigates structural hallucination issues in low-data regions common to deep learning models. The 2023 upgraded version, RoseTTAFold2[21], integrates key AlphaFold2 features such as the frame-aligned point error (FAPE) loss function and cyclic training mechanisms. One year later, the sequence-space diffusion model—Protein Generator (PG)—was publicly released. PG employs iterative noise-based sequence optimization to generate sequence-structure combinations meeting specific functional requirements (e.g., extreme environmental stability, multivalent binding), successfully engineering functional proteins containing non-canonical amino acids[22].

In practical applications, the RoseTTAFold series demonstrates multi-scenario adaptability[23], where RoseTTAFoldNA achieves precise modeling of nucleic acid-protein complex interface[24], while RFjoint enhances protein-protein interaction prediction accuracy through joint optimization strategies[25]. Experimental validations demonstrate that this system attains industry-leading performance in tasks including antibody complementarity-determining region (CDR) modeling[26], prediction of post-translational modification occurrence in GFP-like protein[27], and crystallographic phase determination[28].

Currently, RoseTTAFold and AlphaFold3 represent two dominant technical paradigms in protein design: the former enhances functional design through physics-constrained mechanisms, while the latter leverages data-driven approaches for complex structure prediction[29],[30]. Notably, although RoseTTAFold's energy optimization framework demands greater computational resources, its physical constraint mechanisms provide unique advantages for functional protein design[31],[32]. Recent studies demonstrate that hybrid architectures combining both paradigms enhance design success rates, charting a developmental path for protein engineering platforms that balance physical plausibility and structural novelty[33]–[35].

## 2. Protein Design and Sequence Optimization

### 2.1 Diffusion Generative Models – RFDiffusion

Denoising Diffusion Probabilistic Models (DDPMs) [36], as powerful generative artificial intelligence models, have demonstrated exceptional performance in image and language generation domains. These models iteratively recover clean data from random noise by simulating and learning forward diffusion processes and reverse denoising processes. The Baker Lab developed the RFDiffusion model based on SE(3)-Transformer through deep integration with protein structure prediction networks[37]. By simulating forward diffusion and reverse denoising processes, this model achieves iterative generation of three-dimensional protein structures conforming to geometric constraints and chemical rules from random noise, successfully applied to complex scenarios including functional protein design and construction of bioactive peptide cages.

As it is shown in Figure 4, the training process of RFDiffusion comprises two core stages: feature perturbation and structural reconstruction, corresponding to the forward diffusion and reverse denoising processes. The forward diffusion phase employs a bimodal noise injection strategy: At the local scale, the residue gas mechanism progressively applies independent Gaussian noise to rotation-translation matrices. The probability density function exhibits a bell-shaped curve with a distinct mean and standard deviation. At the global scale, the ISO(3) diffusion model introduces geometric perturbations to three-dimensional conformations. This noise is not simply random points, but rather systematically distorts and deforms the protein's three-dimensional spatial structure by injecting noise into rotation matrices and displacement vectors, thereby altering the overall data architecture and morphological features. This multi-scale noise scheduling strategy simulates the gradual transition of protein structures from ordered to disordered states. Gaussian noise predominantly perturbs the independent dimensions of atomic coordinates, while ISO(3) noise reshapes the global topological architecture through stochastic transformations of rotation matrices. The reverse denoising phase employs an SE(3)-Transformer architecture to process three-dimensional spatial information, progressively restoring the target conformation through iterative prediction of noise distributions. The generative process at each timestep can be formalized as follows: Noisy data $X_t$ is processed through RoseTTAFold to extract latent features and predict an approximate backbone fold conformation $X_0$, though this prediction remains substantially divergent from the true structure. Subsequently, the intermediate state $X_{t-1}$ is computed using the interpolation function interp($X_t$, $X_0$) + ε and output as the input for the next timestep cycle. Notably, the model incorporates AlphaFold2's iterative training mechanism, employing a Self-conditioning strategy that takes both the current timestep's noisy data $X_t$ and historical predictions $X_0$ as inputs. This approach enables the model to fully capture data dynamics and significantly improves trajectory stability during the denoising process. The model establishes a joint loss function that coordinates the optimization of mean squared error (MSE) and Kullback-Leibler (KL) divergence: the former constrains the statistical distribution discrepancy between predicted and actual noise, while the latter ensures probability distribution alignment during the generative process. This dual optimization mechanism enables the system to effectively achieve noise prediction and structural refinement through gradient descent in parameter space, even in the absence of true structure supervision, significantly improving the physical plausibility of generated conformations.

Upon its initial release, RFDiffusion demonstrated exceptional cross-scenario design capabilities. By employing a conditional guidance strategy that overcomes the geometric constraints of traditional methods, it exhibited enhanced adaptive design capabilities compared to other generative models after task-specific fine-tuning, successfully achieving de novo design of neutralizing snake venom toxin proteins[38] and inhibitors targeting the α-helix domain of H3 protein[39]. In the same year, the RFDiffusion All-Atom (AA) version was released [40], which, through transfer learning strategies, adapted RoseTTAFold's pre-trained weights to diffusion tasks, enabling precise construction of small molecule ligand-binding pockets. This advancement maintains backbone generation capability while simultaneously optimizing side-chain conformations and ligand interactions, establishing a new paradigm for structure-based drug design.

**2.2 Structure-Guided Sequence Optimization—ProteinMPNN**

In the field of protein structure prediction and design, most models typically take known amino acid sequences as input to predict their corresponding three-dimensional structures. The inverse folding technique overcomes the dependency limitations on natural templates by inversely deducing the optimal amino acid sequence for a target three-dimensional structure, thereby enabling the design of novel artificial proteins not found in nature and opening innovative pathways for solving practical problems. Among these developments, ProteinMPNN, developed by the Baker team in 2022, stands as one of the representative deep learning models in this field[41]–[43].

ProteinMPNN employs an enhanced graph neural network (GNN) architecture[44] to achieve sequence-structure co-optimization. As it is shown in Figure 5, ProteinMPNN constructs a message-passing neural network (MPNN) [45]–[47]through an encoder-decoder module with three 128-dimensional hidden layers. Innovatively encoding spatial distance features of backbone atoms (N, C, O, $C_α$, and $C_β$) as geometric constraint information for nodes and edges, this feature encoding strategy demonstrates superior residue interaction capture capability compared to traditional methods relying on backbone dihedral angles and rotational orientation extraction, providing stronger inductive bias for sequence design in complex topological structures. Nodes iteratively aggregate neighborhood feature information through the message-passing mechanism, dynamically updating their representation vectors until convergence. The decoding layer of ProteinMPNN adopts a sequence-agnostic stochastic decoding strategy, which maximizes the utilization of sequence contextual information by traversing the global combinatorial space of amino acid permutations. This mechanism overcomes the limitations of traditional fixed-order decoding approaches, allowing sampling initiation from arbitrary residue positions while enhancing both sequence prediction efficiency and the completeness of topological constraint information capture. For the design of symmetric proteins and repetitive structures, the stochastic decoding strategy significantly improves the design efficiency and geometric precision of complex topological proteins through coupled co-optimization of intra-chain or inter-chain equivalent residues, combined with a dynamic locking mechanism for critical structural regions. To address the prevalent backbone coordinate deviation issues in experimental structure determination, the research team introduced multi-scale Gaussian noise perturbations during the training phase, enhancing the model's generalization capability for non-ideal backbone inputs and increasing its tolerance to structural defects, thereby improving robustness. Furthermore, through controlled gradient temperature modulation, the diversity of designed sequences has been increased to 4.7 times that of conventional methods while maintaining the conservation of core functional sites (>95%), better satisfying the function-stability trade-off requirements in practical applications.

The breakthrough application of ProteinMPNN lies in its capability for redesigning complex functional proteins. The research team successfully rescued targets where Rosetta-based design had failed through re-optimization, achieving an increased proportion of interfacial polar residues that significantly improved the in vitro assembly efficiency of nanomaterials. To date, ProteinMPNN has expanded into multiple domains, including enzyme function optimization, immune proteins[48], monomeric and transmembrane proteins[49],[50], and protein complexes[51]. For instance, by constraining key residues to generate functionally specific sequences, it has refined the functional classification system of hydrolases, establishing an interpretable sequence-function mapping framework for multifunctional enzyme engineering[52]. In the design of non-heme iron enzymes, the strategy combining the fixation of key functional sites with stability optimization provides a scalable framework for enzyme design[53]. Recent studies further integrate ESMFold and Rosetta toolkits to systematically optimize catalytic properties through iterative design and evolutionary frameworks, offering a stepwise optimization solution for complex enzymatic functions[54].

Since its inception, ProteinMPNN has demonstrated exceptional versatility and robust functionality in the field of protein design, spawning numerous modular extensions based on its core architecture such as ThermoMPNN[55],[56], FAMPNN[57], LigandMPNN[58], Prot2Chat[59], and the Rosetta toolkit[60]. These extensions feature functional enhancements targeting distinct design objectives (stability, conformation, ligand binding, etc.), collectively advancing multi-scale modeling capabilities in protein design. This fully demonstrates the extensibility potential of the ProteinMPNN architecture and its promising applications in biotechnology and medical fields.

## 3. Multi-Model Applications

With the deepening application of deep learning models in protein engineering, collaborative design strategies integrating multi-model advantages have become a critical pathway to overcome bottlenecks in complex biomolecular design. In this context, prediction models such as AlphaFold are frequently employed as design methodologies for novel proteins and evaluation standards for new model development[61]–[63]. This paradigm leverages deep learning technologies to construct models encompassing cascaded geometric generation, sequence optimization, and structural validation modules, as exemplified by the representative methods mentioned above. In recent years, researchers have successfully implemented cross-domain designs using this paradigm as the primary approach, spanning chaperone proteins, nanomaterials, enzymes[64], antibodies[65],[66], and sensors[67]–[69], significantly enhancing the design efficiency and success rates of functional proteins.

### 3.1 Binder Design

Binder proteins play pivotal roles in cellular homeostasis regulation and disease treatment through their capacity for specific recognition of proteins, nucleic acids, and small molecules. While traditional computational methods have achieved several innovative designs[70]–[72]—such as NeoNectin for modulating integrin α5β1 in regenerative medicine and next-generation therapeutics[73], and proteins accommodating excited-state coupling of chlorophyll special pairs[74]—their limitations persist due to inherent template dependency and time-consuming manual optimization processes, which hinder the balance between binding affinity and specificity. Current applications of deep learning-based models have substantially overcome these challenges[75]–[78]. As it is shown in Figure 6, the three-stage workflow, established through the integration of geometric generation, sequence optimization, and structural validation modules, facilitates experimental validation and screening to yield designs aligned with predefined objectives[79],[80]. A representative implementation of this workflow employs RFDiffusion to generate geometrically complementary topological frameworks for binders, followed by ProteinMPNN-driven optimization of interfacial residue sequences to enhance binding affinity, and culminates in AlphaFold-mediated prediction of binding modes coupled with candidate molecule screening.

In 2024, the Baker group achieved a series of breakthroughs in ligand design for dynamic protein systems by refining the RFDiffusion framework. To address the challenge of intrinsically disordered proteins (IDPs/IDRs) lacking stable three-dimensional structures, researchers introduced a "target secondary structure specification" function combined with a "dual partial denoising" strategy. This approach enabled coordinated sampling of target-ligand conformational space, significantly enhancing the exploration of binding interface diversity [81] [82]. At the technical implementation level, the model integrated β-strand structural constraints, interfacial edge conditions, and secondary structure adjacency matrix (ADJ) features, successfully generating specific binding interfaces capable of recognizing non-canonical conformations (e.g., twists, bulges) in target β-sheets[83]. In November of the same year, the team extended this methodology to peptide-MHC complex (pMHC) systems. By identifying exposed peptide residues outside the MHC binding groove as hotspots and constructing arc-shaped backbone templates, they designed functional ligand proteins forming high-density contacts with dynamic peptide segments[84], providing a novel methodological framework for immunotherapy. Integrating ProteinMPNN's sequence optimization capabilities with AlphaFold's structural prediction functions, the researchers established a high-efficiency binding site engineering platform, achieving transformative progress in both binding site engineering and functional binder development[85]. In the field of fluorescent protein engineering, ProteinMPNN successfully achieved sequence diversification of CagFbFP variants while maintaining fluorescence properties by fixing 20 key chromophore-interacting residues. This demonstrates its potential for exploring non-natural sequence spaces under functional constraints[86]. The same technological framework was applied to amyloid inhibition design, where researchers constructed protein traps with deep peptide-binding grooves. Combined with surface residue optimization strategies, this approach stabilized disordered peptide segments in β-sheet conformations, achieving nanomolar binding affinity and providing a novel strategy for targeting protein aggregation diseases[87]. In pathogen toxin neutralization, RFdiffusion was employed to generate geometrically complementary backbone structures for the receptor interface of the TcsL toxin. Following ProteinMPNN-driven sequence optimization and AlphaFold-based affinity screening, sub-100 pM inhibitors were developed. These inhibitors effectively prevented toxin-induced pulmonary edema in mouse models, marking the first validation of the in vivo efficacy of a fully computational design strategy[88]. This framework was further extended to develop inhibitors for Clostridioides difficile toxin TcdB[89]. Notably, addressing the design bottleneck for flat targets (e.g., TNFR1), researchers implemented a conditional diffusion-based strategy to construct topologically matched deep interfaces. Through partial diffusion-driven receptor subtype specificity reprogramming, they not only established a record-breaking 10 pM monomeric binder but also pioneered a programmable specificity design paradigm[90].

These studies validate the technical advantages of deep learning toolchains across three dimensions: geometric complementarity design (RFdiffusion), sequence-structure co-optimization (ProteinMPNN), and binding mode validation (AlphaFold). Notably, modular strategies have resolved critical challenges faced by traditional methods, such as complex topological targets and dynamic specificity modulation[91], providing a scalable molecular design platform for infectious disease treatment and immune regulation.

**3.2 Optimization of Nanomaterials**

Natural protein nanomaterials demonstrate significant application value in drug delivery systems and vaccine development due to their self-assembly properties. However, early design methods primarily relied on limited modifications of natural protein structures, constrained by factors such as size and shape, which hindered their ability to meet complex functional demands. The introduction of computational methods like Rosetta enabled atomic-level control over protein subunit interactions. For instance, smart protein fibers responsive to ligand binding[92] or triggered by subtle pH changes for self-assembly were developed[93]. In recent years, the integration of deep learning with traditional physical modeling has further driven technological innovation in this field, successfully constructing pseudo-symmetric multimeric protein materials[94],[95] and complex nanocage architectures[96],[97]. For example, through icosahedral pseudo-symmetric heterotrimeric design, researchers engineered large-scale nanocages comprising 240, 540, and 960 subunits[98].

The deep integration of deep learning models with protein nanomaterial design has significantly advanced breakthroughs in structural flexibility and design efficiency within this field. As it is shown in Figure 7, the collaborative design strategy combining the RDFDiffusion diffusion model and the ProteinMPNN sequence optimization module has evolved into a standardized technical paradigm: the former generates target topological frameworks through three-dimensional conformational space sampling, while the latter performs high-precision sequence design for interface residues. In the field of protein nanocage engineering, ProteinMPNN has effectively enhanced assembly stability and enabled customized designs for specific biological functions (e.g., molecular encapsulation or targeted delivery) by optimizing the interfacial interaction networks of cage structures[99],[100]. Furthermore, the RFjoint inpainting algorithm developed based on the RoseTTAFold architecture, in conjunction with RDFDiffusion, has successfully achieved high-throughput rational design of transmembrane β-barrel proteins and their nanochannels[101]. Particularly noteworthy is the groundbreaking work utilizing RDFDiffusion to generate asymmetric interface topologies and ProteinMPNN to design heterotypically complementary interface sequences, which has enabled the construction of Janus-type protein nanoparticles with precisely controlled geometric morphologies and spatially segregated surface functionalities[102]. This innovative design paradigm transcends the limitations of traditional symmetry-driven assembly principles, providing a novel technological pathway for developing modular intelligent nanomaterials.

In the design of multi-component nanomaterial systems, deep learning models demonstrate exceptional geometric compatibility capabilities and multi-scale assembly control properties. Research teams utilized RFDiffusion to generate hierarchical structural units with linear, curved, and angularized features. By integrating ProteinMPNN to optimize cooperative interaction networks at interface residues, they achieved high-precision programmable assembly of complex nanostructures[35]. Notably, ProteinMPNN employs a polarity residue-dominated sequence optimization strategy through deep analysis of natural protein interface contact patterns, significantly enhancing charge complementarity across component contact surfaces. This design approach effectively reduces hydrophobic surface exposure of protein monomers in unassembled states, successfully constructing a two-component tetrahedral nano assembly with superior thermodynamic stability[103].

In summary, these technological breakthroughs not only overcome the bottleneck of insufficient geometric adaptability in structural modules inherent to traditional methods but also mark a revolutionary paradigm shift in protein nanomaterial design—from empirical trial-and-error to computation-driven approaches. This advancement lays the technical foundation for developing smart nanodevices with customized biological functionalities.

## 4.Discussion

In recent years, deep learning-driven protein engineering has achieved a paradigm shift from structural analysis to functional customization. Algorithms represented by the AlphaFold model, through an iterative architecture combining evolutionary covariance information encoding and diffusion optimization, have elevated the prediction accuracy of single-chain proteins to the level of experimental resolution (RMSD < 1.0Å). Their application boundaries have further expanded to complex systems such as protein-nucleic acid complexes. Meanwhile, RoseTTAFold's three-track neural network synergistically integrates sequence features, residue contact maps, and spatial coordinate constraints. Coupled with the geometry-aware capabilities of the SE(3) Transformer, it significantly reduces physical implausibility in conformational generation. The collaborative innovation of these two model types has given rise to a joint design framework combining RFDiffusion and ProteinMPNN. This framework employs a cascaded strategy where diffusion models generate topological scaffolds and inverse folding algorithms optimize interface residues, accelerating the industrialization of protein design in biomedicine and synthetic biology. Applications such as antibody, enzyme, and nanomaterial design and functional optimization are providing new paradigms for drug delivery system development.

As of March 2025, the latest advancements in deep learning techniques within the field of protein design further underscore their transformative potential. In the realm of nanomaterial design, RFDiffusion has achieved nanostructure regulation based on external variables (e.g., pH or ionic strength) through a conditional generative framework. Recent studies indicate that integrating molecular dynamics (MD) simulations with diffusion models enables the development of environmentally responsive protein materials, such as nanostructures capable of self-assembly or disassembly under specific conditions, offering new possibilities for controlled drug release applications. Additionally, the synergistic optimization of RFjoint and ProteinMPNN has enabled the precise design of transmembrane β-barrel nanopores, with pore size tunability supporting the development of ion-selective sensors and artificial cell membrane systems. In the area of binder protein design, AlphaFold3, through architectural improvements, has significantly enhanced the prediction accuracy of protein-protein interfaces (PPIs) and intrinsically disordered proteins (IDPs). For instance, binder designs targeting specific molecules (e.g., EGFR) using RFDiffusion and ProteinMPNN have demonstrated high affinity and potential therapeutic efficacy in experimental validation. For conformational predictions of IDPs, conditional diffusion models have optimized the capture of dynamic conformations, providing novel strategies for intervening in targets related to neurodegenerative diseases. Building on the rapid development of these technologies, multimodal data integration has emerged as a key trend driving progress in this field. The integration of RoseTTAFoldNA with cryo-electron microscopy (cryo-EM) density maps has elevated the prediction accuracy of nucleic acid-protein complexes to an RMSD < 0.4 Å. This approach has been applied to resolve virus nucleic acid-protein interactions, supporting structural optimization of mRNA vaccines. Meanwhile, the AlphaFold Protein Structure Database (AlphaFold DB), through continuous upgrades and the addition of user dataset submission capabilities, has further facilitated the realization of customized protein design.

Despite remarkable achievements, this technological field still faces limitations and unresolved challenges. While models like RFDiffusion can efficiently generate topological structures in nanomaterial design, their performance is constrained by the complex conditions of application scenarios (e.g., in vivo solvent environments and protein-protein interactions), leading to reduced efficacy. In binder protein design, most models remain immature in modeling interactions with small molecules and non-classical ligands (e.g., metal ions). Although progress has been made in predicting dynamic conformations, comprehensively capturing functionally relevant states remains challenging, resulting in designed proteins with insufficient stability in practical applications. Furthermore, the heavy reliance of these models on standardized structural data from the Protein Data Bank (PDB) represents a significant bottleneck. Membrane proteins, small molecule interactions, and non-natural proteins are underrepresented in the PDB—for instance, membrane proteins constitute only a small fraction of the database entries, limiting the model's generalization capabilities in these areas. The low throughput of experimental validation exacerbates this issue. In the AlphaFold Protein Structure Database (AlphaFold DB), only a subset of predicted structures has been validated through wet-lab experiments, and high-affinity binding does not always translate into expected functionality.

For example, nonspecific binding observed in cellular experiments has constrained their potential for clinical translation.

These limitations stem from multiple constraints related to technology and data. On the technical front, current models such as AlphaFold and RoseTTAFold are primarily optimized for static structure prediction, whereas protein folding and function in biological environments involve dynamic processes spanning milliseconds to seconds. Existing attention mechanisms and diffusion models struggle to effectively simulate these timescales. In terms of validating and utilizing design outcomes, the workflow between model design and experimental validation remains disjointed. Additionally, the immaturity of screening techniques results in lengthy and costly validation cycles. Regarding training data, the inadequacy in modeling small molecule interactions is closely tied to the limited chemical diversity of PDB training data. Small molecule data in the database are predominantly focused on common ligands (e.g., GTP, heme), with insufficient capacity to infer chemical rules for rare or synthetic ligands, leading to prediction biases in non-standard scenarios. Furthermore, although the AlphaFold DB covers a vast number of sequences, its structural diversity still falls short of fully representing the complexity of natural proteins, particularly in disordered regions and membrane protein domains, where further exploration is needed. A deeper issue lies in the scarcity of negative sample data. Model training data predominantly consists of successful cases, lacking systematic mappings of failed structure-function relationships, which hampers the models' ability to learn from errors.

To address these challenges, future research should integrate advanced modeling and data strategies while exploring diverse application prospects. By combining multiphysics models with diffusion models, intelligent nanostructures adaptable to complex in vivo environments can be designed, while leveraging reinforcement learning and molecular dynamics simulations to optimize the dynamic prediction of intrinsically disordered proteins (IDPs) and protein-protein interfaces (PPIs) and to enhance non-classical ligand interactions. To tackle issues of data scarcity and algorithmic limitations, developing lightweight models that fuse multimodal inputs—such as cryo-electron microscopy density maps and small-molecule microenvironment data—can improve temporal sampling capabilities. Alternatively, employing generative adversarial networks and small-molecule fragment libraries to enrich the derivation of chemical rules offers another approach. These advancements will propel the dynamic regulation design of metabolic pathways based on diffusion models in synthetic biology. Furthermore, through multiphysics analysis of protein-material interfaces, the development of intelligent nanocarriers and the design of chaperone proteins—improved for potential applications in cancer immunotherapy and neurodegenerative disease intervention—will foster synergistic progress in biomaterial design and precision medicine research.

## 5. Conclusion

Deep learning is redefining protein engineering through intelligent closed-loop integration of structure, function, and design. This review identifies three transformative trends: 1) The fusion of generative models with physical constraints transcends geometric limitations, enabling functional customization across scales; 2) Cooperative modeling of dynamic conformations expands predictive capabilities for challenging targets like membrane proteins and disordered regions; 3) Multimodal learning frameworks unify the handling of complex biomolecular interactions. These advancements have not only yielded breakthrough applications such as atomically precise nanopores and picomolar binders but also redefined computational biology's role in synthetic life systems.

The fundamental challenge lies in the inherent tension between biological complexity and computational abstraction: Static structural datasets inadequately represent physiological conformational dynamics, while incomplete energy functions introduce systematic biases in non-canonical interaction prediction. The solution pathway involves constructing a reinforcement learning ecosystem integrating generation, simulation, and validation, where molecular dynamics provide temporal resolution, cryo-EM delivers spatial precision, and generative models contribute creative potential. This evolving fusion intelligence may catalyze a paradigm shift from protein design to the rational construction of living systems. As the field progresses, the seamless integration of probabilistic generation and deterministic biophysical rules will be crucial for bridging the gap between digital designs and functional realities in cellular environments.

## 6. Author Acknowledgments

Y.W. Wang thank Zhejiang Provincial Natural Science Foundation General Project （LY23A40002）； Wenzhou Science and Technology Plan Project（L20240004）；
Y. Wang thank the Research Fund of Wenzhou Institute, UCAS (WIUCASQD2021043);

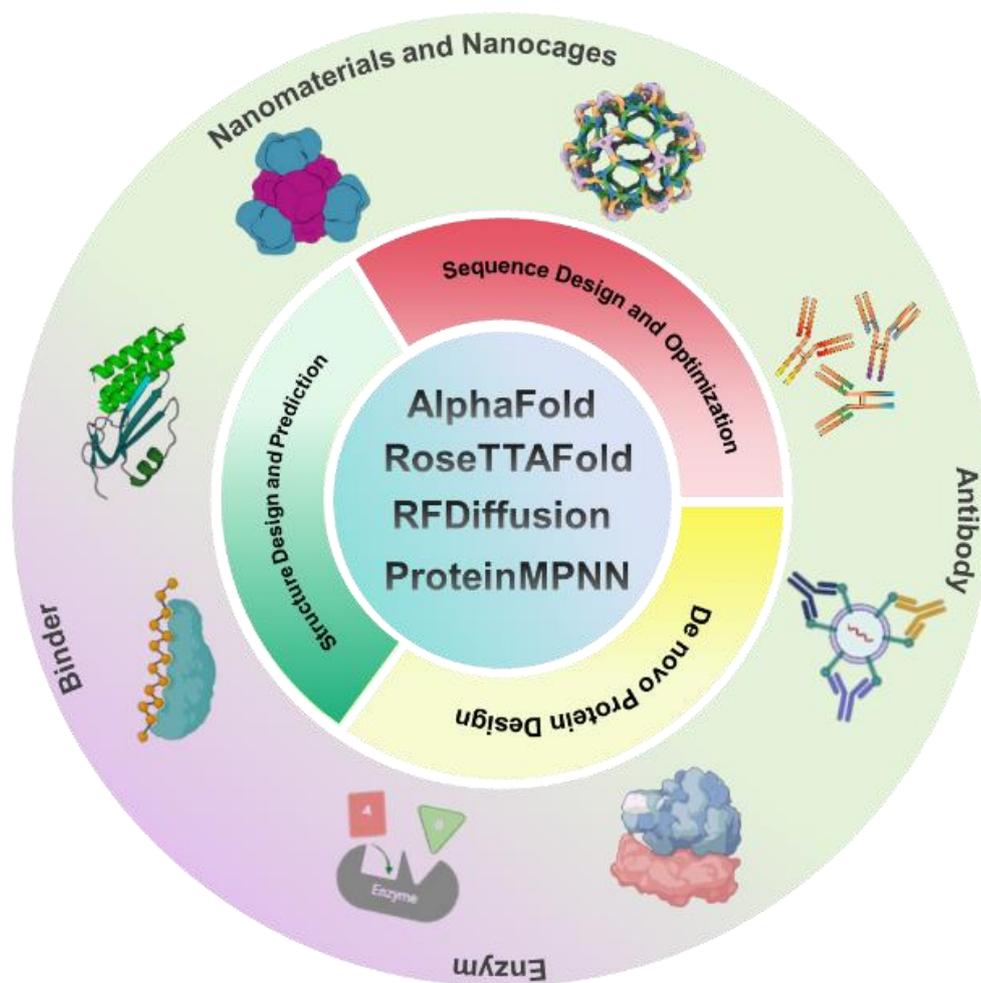

**Figure 1. Model-driven protein engineering development based on deep learning technology.** This figure systematically presents the core model architecture and its co-design paradigm of deep learning technology in protein engineering through hierarchical concentric circles. Beginning with the foundational models discussed in this paper — AlphaFold, RoseTTAFold, RFDiffusion, and ProteinMPNN — the framework encompasses protein structure prediction and design, sequence optimization, and De novo protein design. Furthermore, it illustrates the synergistic integration of multiple models for interdisciplinary applications under emerging development trends, including nanomaterials, protein binders, antibodies, and enzyme engineering.

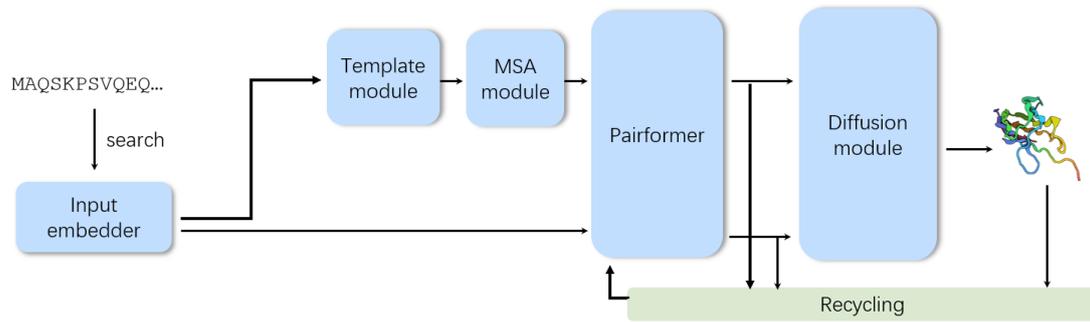

**Figure 2. Multi-module collaborative framework for structure prediction in AlphaFold3.** The architecture of AlphaFold3 builds upon AlphaFold2 with critical enhancements. Input sequences are initially processed through a search module and an input embedding layer. The encoded information sequentially traverses the template module and multiple sequence alignment (MSA) module, followed by integration in the Pairformer module for residue-pair feature refinement. This processed information is subsequently fed into a diffusion-based structure decoder to ultimately generate the protein's three-dimensional conformation.

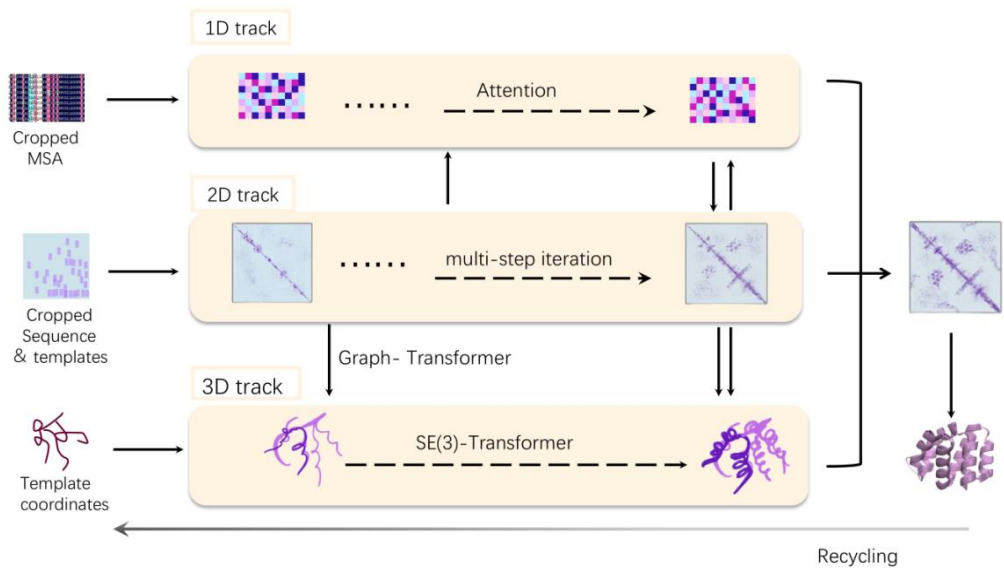

**Figure 3. Three-track architecture of RoseTTAFold for multimodal protein structure prediction.** The framework enables simultaneous processing of sequence features (1D), inter-residue distance/orientation matrices (2D), and spatial coordinate embeddings (3D). A unified deep neural network architecture jointly optimizes these multimodal representations through geometric transformation layers and iterative refinement via self-attention mechanisms, ultimately generating high-accuracy three-dimensional protein structural predictions.

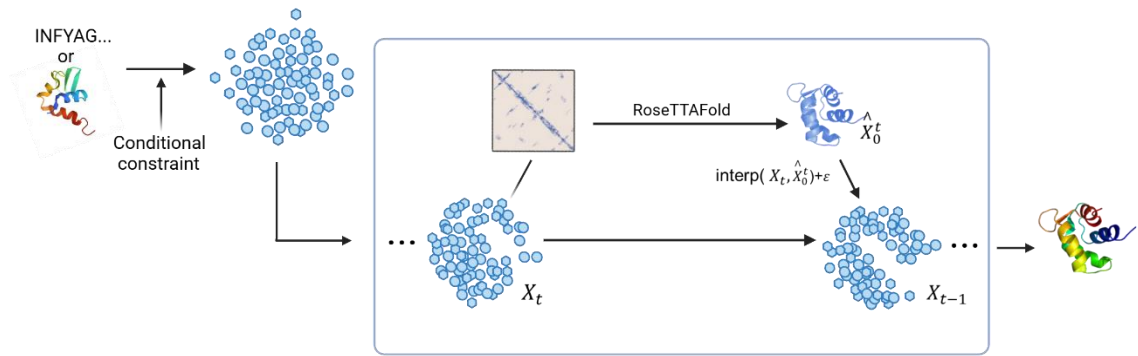

**Figure 4. Architectural framework of RFDiffusion.** The model integrates the SE(3)-Transformer architecture with RoseTTAFold's pre-trained network and Denoising Diffusion Probabilistic Models (DDPMs). This hybrid approach implements a two-phase process: (1) forward diffusion to progressively perturb structural coordinates, followed by (2) iterative backward denoising through SE(3)-equivariant transformations. The DDPM-driven pipeline enables progressive refinement of three-dimensional protein conformations, achieving convergence through geometrically constrained latent space optimization.

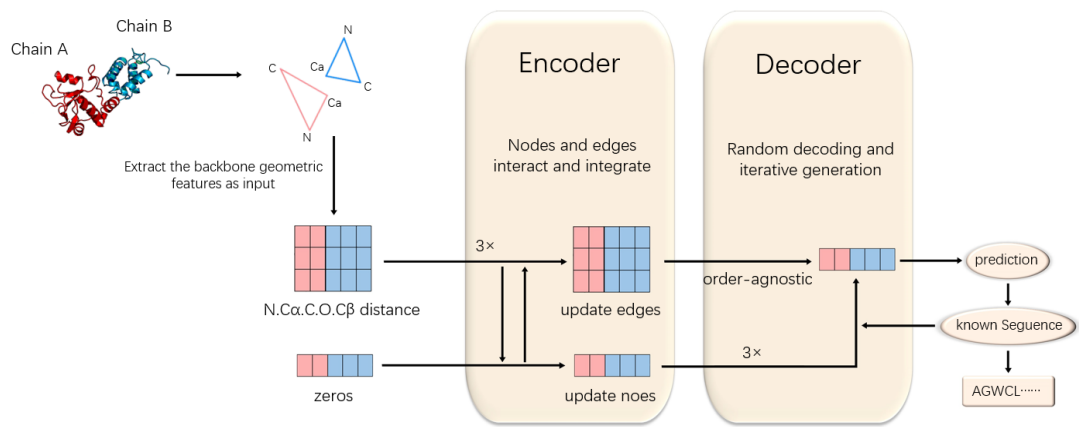

**Figure 5. ProteinMPNN sequence-structure co-design framework.** The architecture extends message-passing neural networks (MPNNs) through a bidirectional graph encoder-decoder system engineered to enforce backbone geometric constraints. Implementing a 3-layer geometric encoder with SE(3)-invariant edge features, autoregressive stochastic decoding with temperature-annealed sampling, and hierarchical residue interaction modeling via 128-dimensional hidden states, this framework achieves iterative sequence-structure co-optimization across three decoder layers.

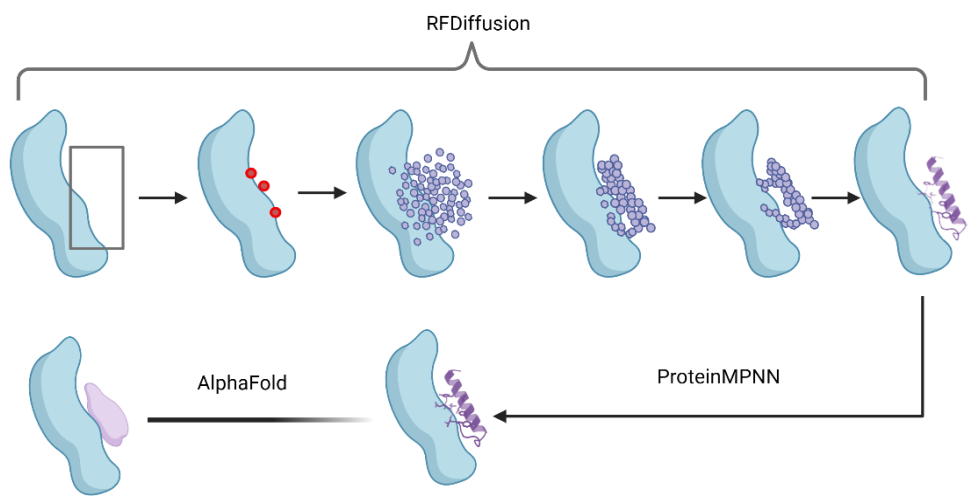

**Figure 6. Multi-model co-design pipeline for high-affinity protein binder development.** This integrative framework synergizes three core components: de novo structural sampling via RFDiffusion geometry-aware diffusion process, binding interface optimization using ProteinMPNN sequence-structure energy landscapes, and structural validity screening through AlphaFold-based folding confidence metrics. The baseline workflow is augmented with specialized modules for dynamic targets, including a dual-stage denoising protocol to enhance conformational diversity and secondary structure priors for disordered protein interfaces. Experimental validation demonstrates nanomolar-scale binding affinities in designed inhibitors for toxin neutralization and peptide-MHC complexes, establishing a robust platform for therapeutic applications ranging from immunotherapy to precision drug discovery.

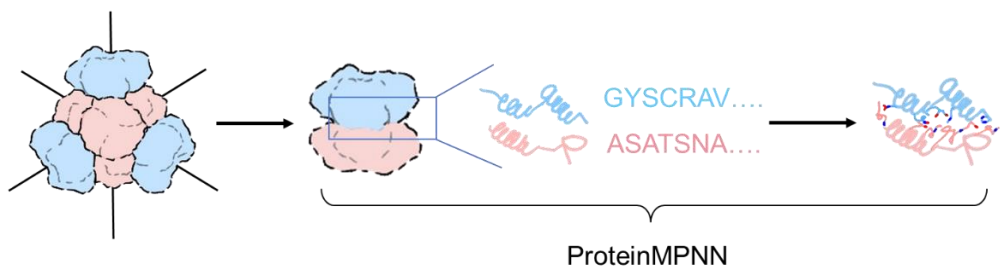

**Figure 7. Deep learning-driven de novo design of protein nanomaterials.** The computational framework employs ProteinMPNN to engineer two-component tetrahedral nanoparticles through interface-specific sequence sampling and side-chain conformation optimization. By strategically enriching polar residues at heterocomponent interfaces (e.g., glutamine/asparagine clusters), the design minimizes hydrophobic surface exposure on monomeric subunits, thereby enabling symmetry-constrained self-assembly in vitro with >90% yield efficiency. This energy landscape-driven approach demonstrates programmable control over supramolecular architecture while maintaining biocompatibility—key for applications in targeted drug delivery and synthetic vaccines.

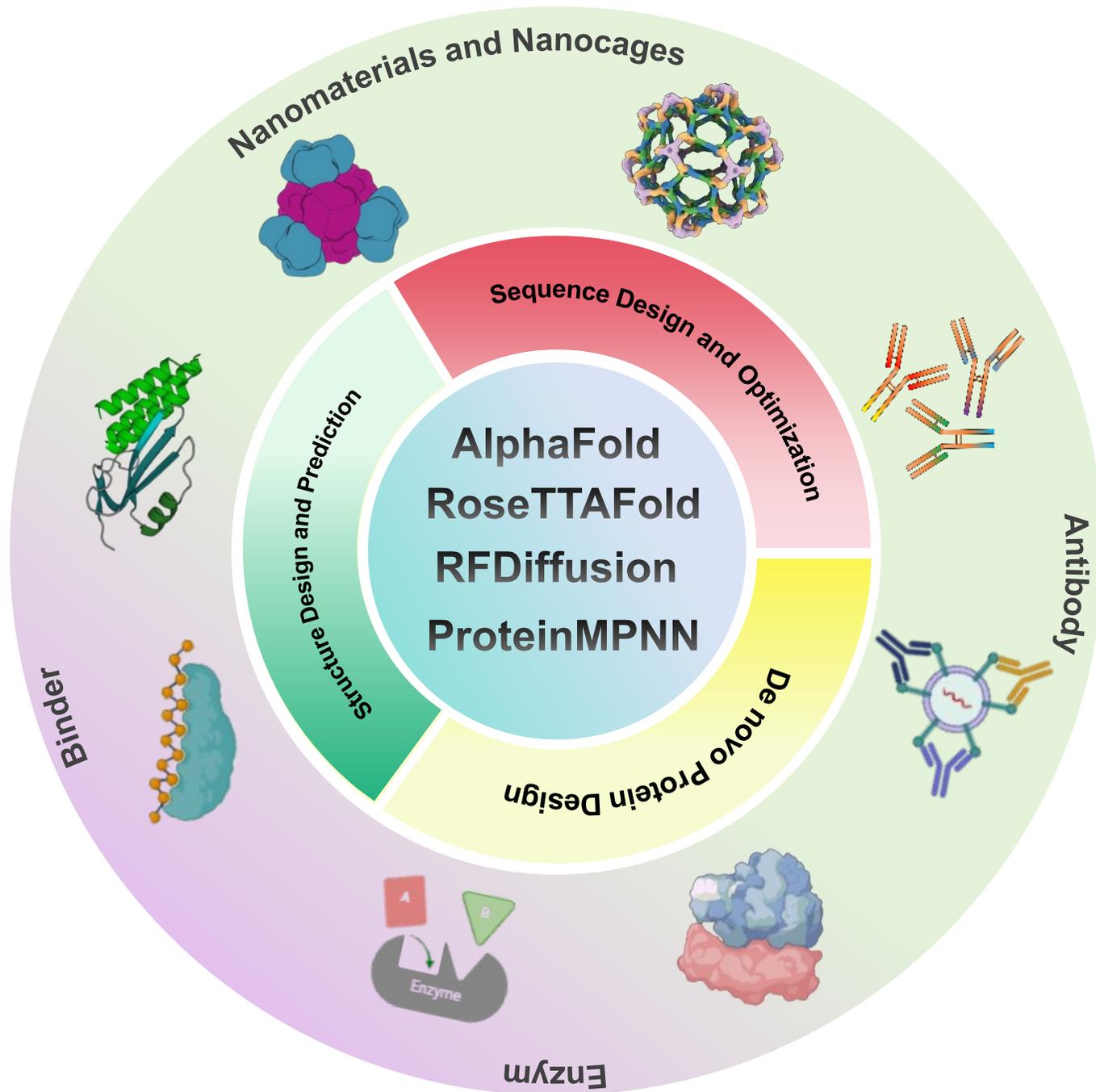

**Figure 1. Model-driven protein engineering development based on deep learning technology.** This figure systematically presents the core model architecture and its co-design paradigm of deep learning technology in protein engineering through hierarchical concentric circles. Beginning with the foundational models discussed in this paper—AlphaFold, RoseTTAFold, RFDiffusion, and ProteinMPNN—the framework encompasses protein structure prediction and design, sequence optimization, and De novo protein design. Furthermore, it illustrates the synergistic integration of multiple models for interdisciplinary applications under emerging development trends, including nanomaterials, protein binders, antibodies, and enzyme engineering.

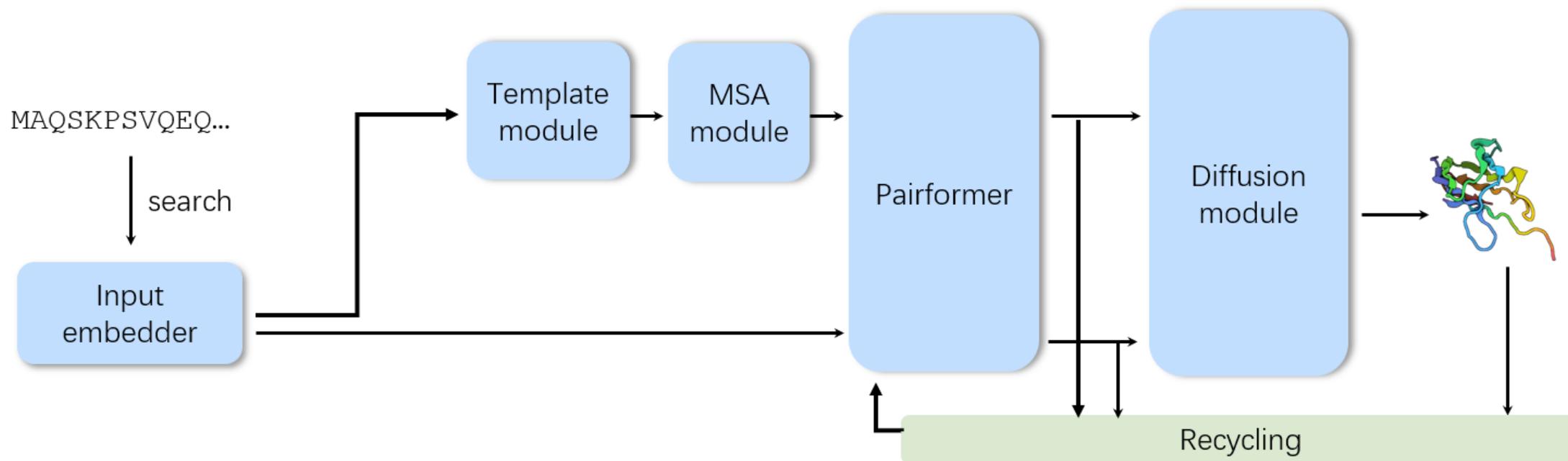

**Figure 2. Multi-module collaborative framework for structure prediction in AlphaFold3.** The architecture of AlphaFold3 builds upon AlphaFold2 with critical enhancements. Input sequences are initially processed through a search module and an input embedding layer. The encoded information sequentially traverses the template module and multiple sequence alignment (MSA) module, followed by integration in the Pairformer module for residue-pair feature refinement. This processed information is subsequently fed into a diffusion-based structure decoder to ultimately generate the protein's three-dimensional conformation.

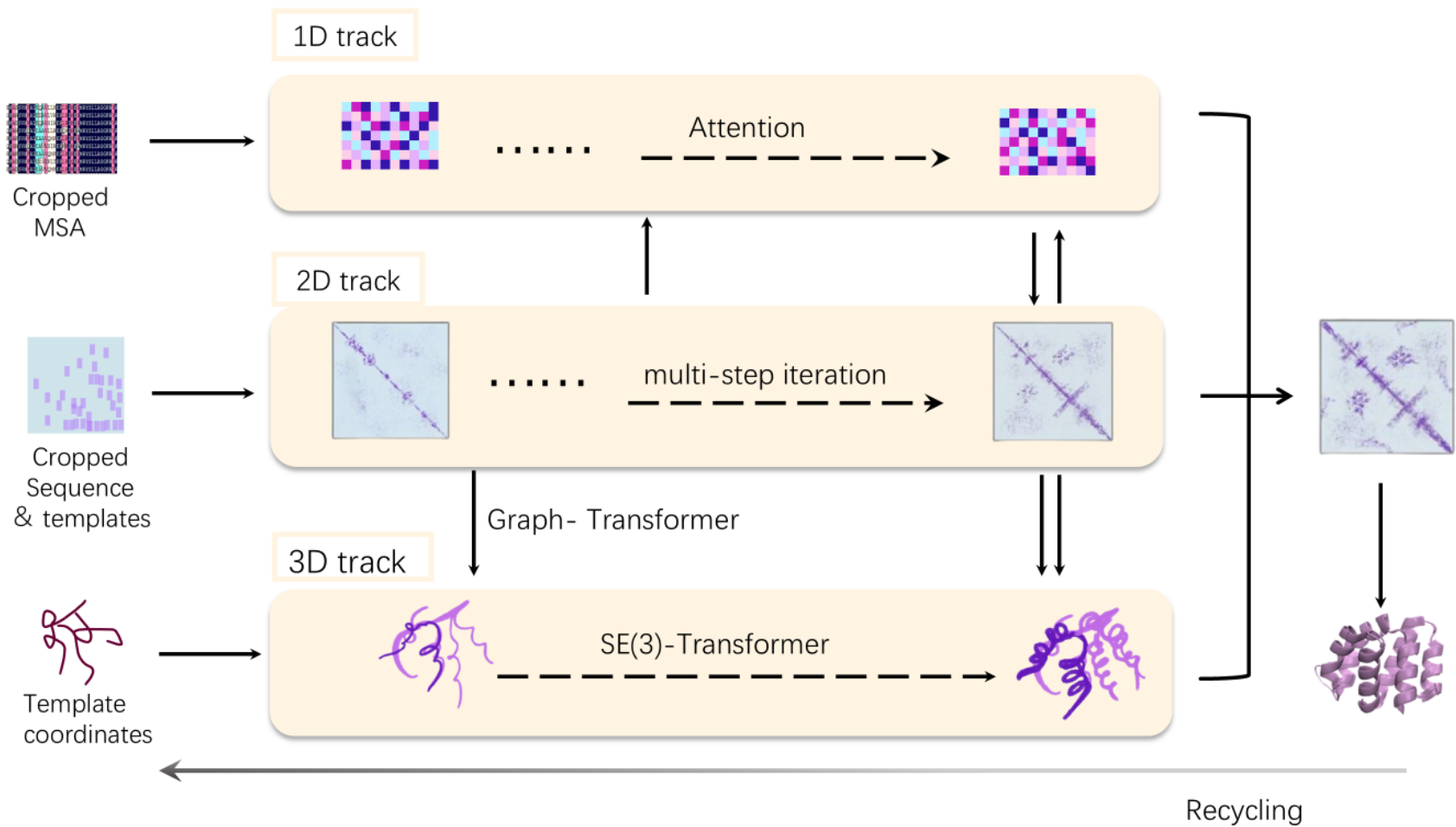

**Figure 3. Three-track architecture of RoseTTAFold for multimodal protein structure prediction.** The framework enables simultaneous processing of sequence features (1D), inter-residue distance/orientation matrices (2D), and spatial coordinate embeddings (3D). A unified deep neural network architecture jointly optimizes these multimodal representations through geometric transformation layers and iterative refinement via self-attention mechanisms, ultimately generating high-accuracy three-dimensional protein structural predictions.

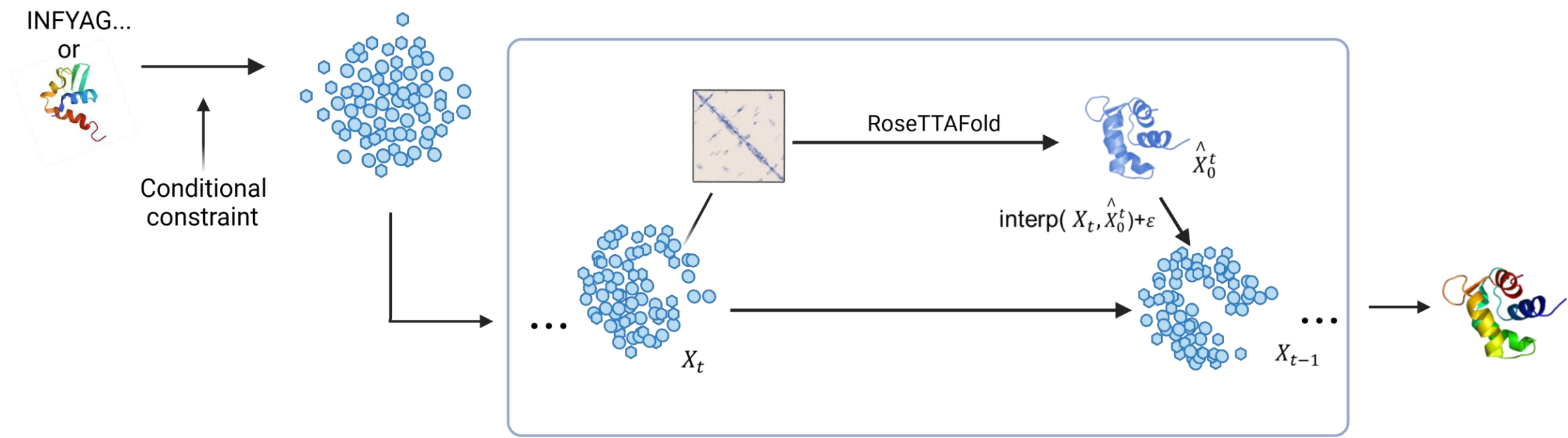

**Figure 4. Architectural framework of RFDiffusion.** The model integrates the SE(3)-Transformer architecture with RoseTTAFold's pre-trained network and Denoising Diffusion Probabilistic Models (DDPMs). This hybrid approach implements a two-phase process: (1) forward diffusion to progressively perturb structural coordinates, followed by (2) iterative backward denoising through SE(3)-equivariant transformations. The DDPM-driven pipeline enables progressive refinement of three-dimensional protein conformations, achieving convergence through geometrically constrained latent space optimization.

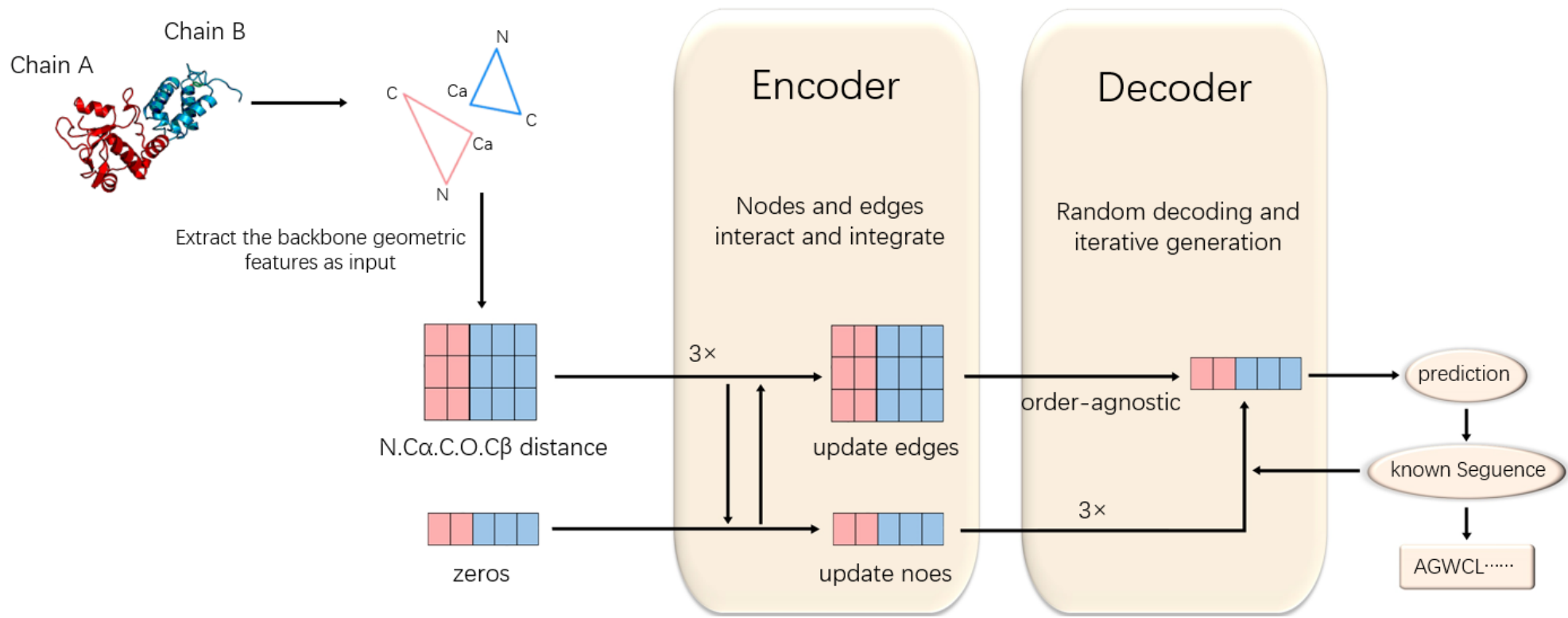

**Figure 5. ProteinMPNN sequence-structure co-design framework.** The architecture extends message-passing neural networks (MPNNs) through a bidirectional graph encoder-decoder system engineered to enforce backbone geometric constraints. Implementing a 3-layer geometric encoder with SE(3)-invariant edge features, autoregressive stochastic decoding with temperature-annealed sampling, and hierarchical residue interaction modeling via 128-dimensional hidden states, this framework achieves iterative sequence-structure co-optimization across three decoder layers.

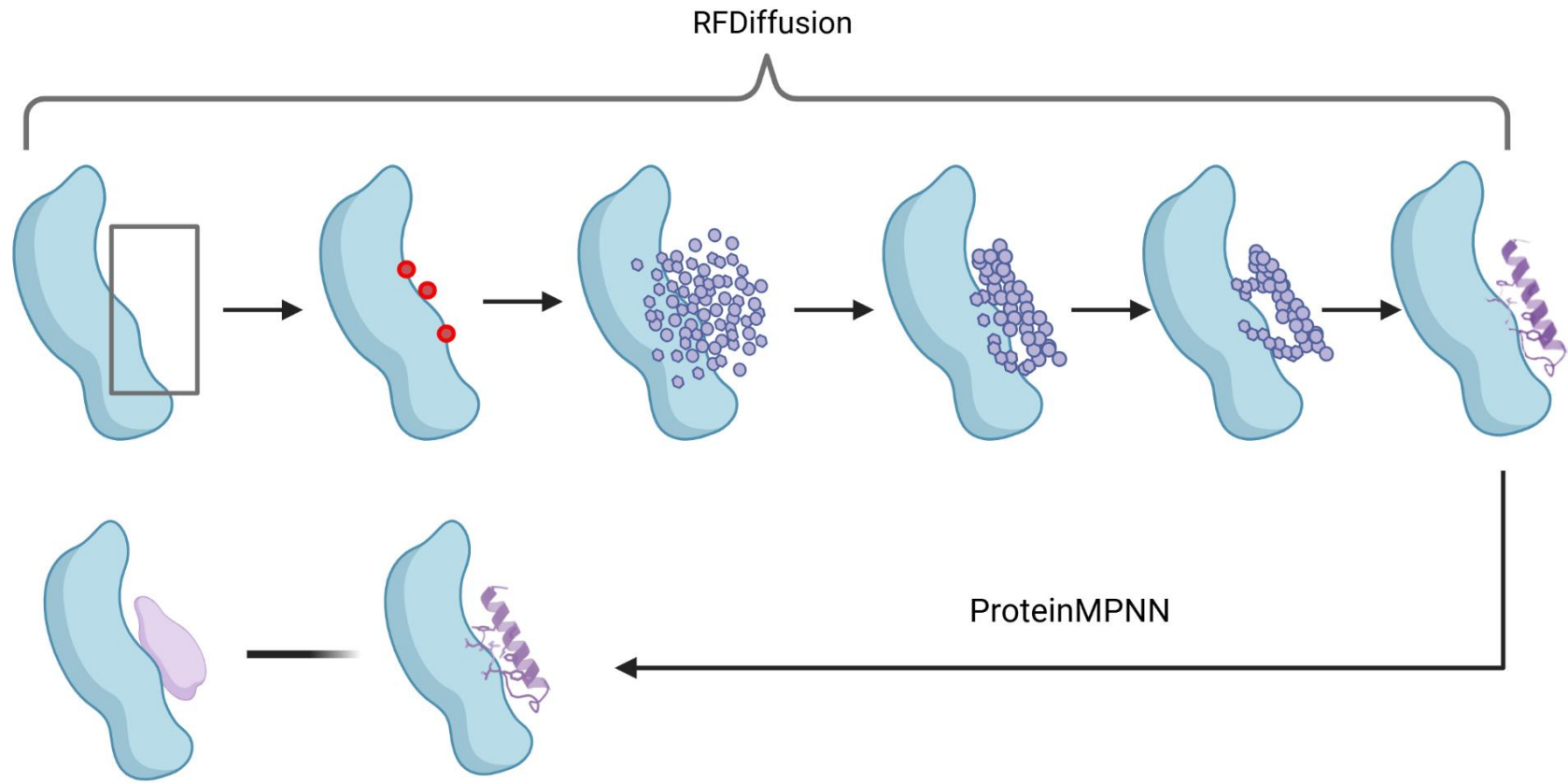

**Figure 6. Multi-model co-design pipeline for high-affinity protein binder development.** This integrative framework synergizes three core components: de novo structural sampling via RFDiffusion geometry-aware diffusion process, binding interface optimization using ProteinMPNN sequence-structure energy landscapes, and structural validity screening through AlphaFold-based folding confidence metrics. The baseline workflow is augmented with specialized modules for dynamic targets, including a dual-stage denoising protocol to enhance conformational diversity and secondary structure priors for disordered protein interfaces. Experimental validation demonstrates nanomolar-scale binding affinities in designed inhibitors for toxin neutralization and peptide-MHC complexes, establishing a robust platform for therapeutic applications ranging from immunotherapy to precision drug discovery.

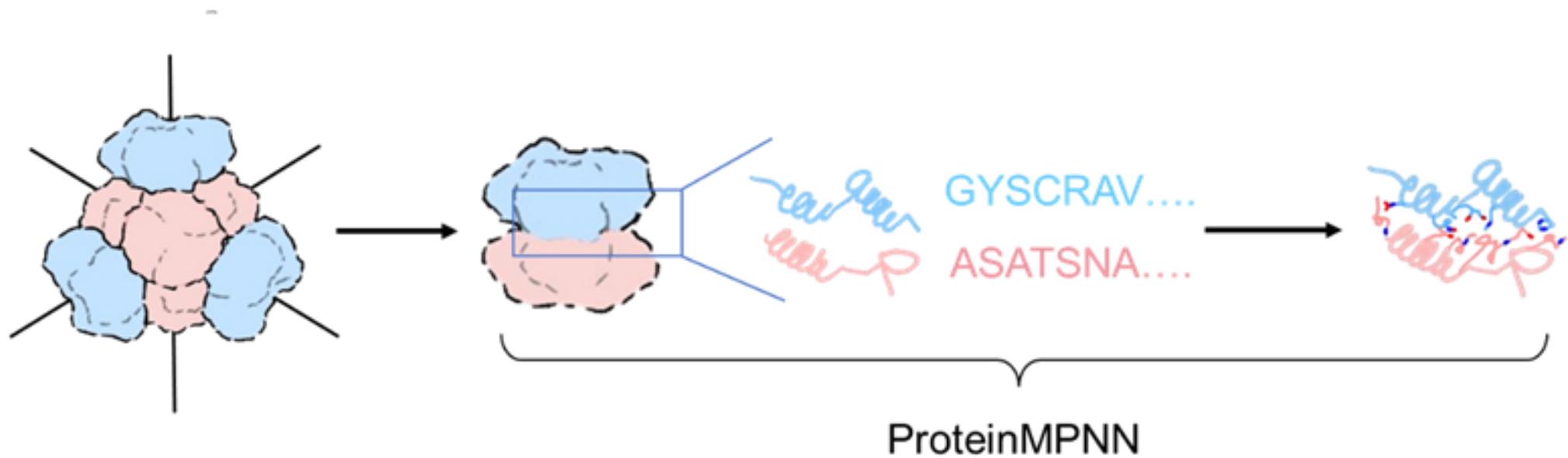

**Figure 7. Deep learning-driven de novo design of protein nanomaterials.** The computational framework employs ProteinMPNN to engineer two-component tetrahedral nanoparticles through interface-specific sequence sampling and side-chain conformation optimization. By strategically enriching polar residues at heterocomponent interfaces (e.g., glutamine/asparagine clusters), the design minimizes hydrophobic surface exposure on monomeric subunits, thereby enabling symmetry-constrained self-assembly in vitro with >90% yield efficiency. This energy landscape-driven approach demonstrates programmable control over supramolecular architecture while maintaining biocompatibility—key for applications in targeted drug delivery and synthetic vaccines.